\documentclass[prd,aps,twocolumn,showpacs,preprintnumbers,amsmath,amssymb,
nofootinbib]{revtex4}
\usepackage[mathscr]{eucal}

\usepackage{graphicx}
\usepackage{dcolumn}
\usepackage{bm}
\begin{document}

\title{Evolution of Metric Perturbations in Quintom Bounce model}

\author{Yi-Fu Cai and Xinmin Zhang}

\affiliation{ Institute of High Energy Physics, Chinese Academy of
Sciences,  P.O.Box 918-4, Beijing 100049, P.R.China}

\begin{abstract}

We in the paper study the metric perturbations generated in a
bouncing universe driven by the Quintom matter. Firstly, we review
the background evolution of Quintom Bounce and the power spectrum of
scalar perturbations. Secondly, we study the non-Gaussianity of
curvature perturbations and then calculate the tensor perturbations
of the model.

\end{abstract}


\maketitle


\section{Introduction}

Inflation was invented to resolve problems existing in hot Big
Bang cosmology, such as flatness, horizon, primordial monopole
problem\cite{Guth:1980zm} (for some early attempts see Refs.
\cite{Starob}). After more than twenty years' development, one has
obtained a deep sight at this theory. However, it is still puzzled
by an initial singularity which exists in usual inflationary
models\cite{Borde}.

It has been suggested that, bouncing cosmology, which requires our
universe initially experience a contracting phase before the hot
Big Bang expansion, could provide a solution to the problem of the
initial singularity. For years, models of a bouncing universe have
received a lot of attentions and there have been a number of works
on constructing this scenario. For example, there are models with
singular bounce such as the Pre-Big-Bang \cite{PBB} and
cyclic/Ekpyrotic \cite{Ekp}; also in Refs.
\cite{Bojowald:2001xe,Brustein:1997cv,Biswas:2005qr} some
non-singular bounce models were considered where the gravitation
action was modified by higher order corrections; and, see Ref.
\cite{Novello:2008ra} for a recent review on various models of
bouncing cosmology.

Recently, a new bounce model has been proposed \cite{Cai:2007qw},
dubbed Quintom Bounce. In this model, a bouncing universe was
obtained within the standard 4-dimensional
Friedmann-Robertson-Walker (FRW) framework by making use of
Quintom matter\cite{Feng:2004ad}. The key feature of this model is
that, the null energy condition (NEC) has to be violated for a
short while around the bounce point and after that the
equation-of-state (EoS) of our model $w$ is able to transit from
below $-1$ to above $-1$ which makes the universe enter into
normal expanding history.

The model of Quintom Bounce has presented a very interesting
picture of the early universe. Firstly, the investigation of the
cosmological perturbations in this model \cite{Cai:2007zv} has
shown a combination of some aspects found in some recently studied
non-singular bounce models \cite{Abramo,Brandenberger:2007by} and
some others in singular bounce models
\cite{BGGMV,Lyth,Hwang2,Fabio}. Secondly, in a recent study on
Quintom Bounce, the authors of Ref. \cite{Cai:2008qb} have found
that a concrete model of Quintom Bounce are able to provide a
scale-invariant spectrum in ultra-violet regime, and also give
rise to an oscillation signature which could be verified in the
forthcoming astronomical observations.

One important lesson of studying the primordial curvature
perturbations is to investigate its bispectrum which describes the
non-Gaussianity of the power spectrum\cite{Bartolo:2004if}. There
is a number of information coded in this bispectrum, such as
magnitude, shape, sign, and even running. To probe these
signatures, we expect to distinguish various models of the very
early universe. For example, in the usual slow-roll inflationary
model, it is pointed out that non-Gaussianity is negligible due to
the suppression of slow-roll
parameters\cite{Maldacena:2002vr,Acquaviva:2002ud}; however, in
the models of Ekyrotic/cyclic
\cite{Koyama:2007if,Buchbinder:2007at,Lehners:2007wc} or island
universe\cite{Piao:2007cj}, a large non-Gaussianity is predicted.
Observationally, current cosmological
data\cite{Yadav:2007yy,Komatsu:2008hk} is consistent with Gaussian
distribution, however, the forthcoming observations will be
sensitive to the non-Gaussianity with much higher precision, such
as Planck satellite \cite{Planck}. Given the considerations above,
we in this paper calculate the bispectrum of a Quintom Bounce
model. Our results show that non-Gaussianity in this model is
still suppressed by slow-roll parameters, but there is an
interesting oscillation signal on the non-linear parameter and its
maximal value is mildly bigger than that in the usual scenario of
slow-roll inflationary models.

Another important clue to discover the information of the very
early universe is to study the relic gravitational wave background
(GWB) formed by primordial tensor fluctuations. There have been a
number of detectors operating for the signals of primordial GWB,
e.g. Planck \cite{Planck}, Big Bang Observer (BBO) \cite{BBO},
LIGO \cite{Abbott:2004ig}. Moreover, the indirect detection of GWB
attracts a lot of interests of the next generation of CMB
observations, see related analyses \cite{Verde:2005ff,
Smith:2005mm, Boyle:2005ug,TFCR}. The basic mechanism for the
generation of primordial GWB in cosmology has been discussed in
Refs. \cite{Grishchuk:1974ny,Allen:1987bk}. Usually, inflation
theory predicts that the power spectrum of primordial tensor
fluctuations is scale-invariant and its value is roughly equal to
scalar spectrum times a slow-roll parameter defined as
$\epsilon\equiv-\dot
H/H^2$\cite{Starobinsky:1979ty,Stewart:1993bc}. However, this
so-called consistency relation is not necessary to be valid in a
bounce scenario. For example, Ref. \cite{Boyle:2003km} has
investigated the primordial gravitational waves in singular bounce
models and predicted an undetectable GWB. We in this paper study
the relic gravitational waves in Quintom Bounce with
Coleman-Weinberg potential. Interestingly, we find that the power
spectrum of primordial gravitational waves is scale-invariant both
in ultra-violet and infrared regimes but strongly oscillate in the
middle band which is related to the bounce, and also find a sunken
signature on the energy spectrum. These results on one side
support the conclusions of Ref. \cite{Cai:2008qb} in which the
scalar perturbations were discussed, and on the other side provide
another approaches to searching the signals of a bounce.

The outline of this paper is as follows. In Section II, we briefly
review the basic scenario of a Quintom Bounce model with a the
Coleman-Weinberg potential, and the curvature perturbation. In
Section III, we make the calculation of non-Gaussianity in the
model of Quintom Bounce, and find that the non-linear parameter
$f_{NL}$ is oscillating around a small value which coincide with
that in usual inflation model. In Section IV, we investigate the
tensor part of gravitational perturbations in this model, and then
give the power spectrum and spectral index correspondingly.
Considering the influences of possible factors during the late
time evolution, we discuss the transfer function for the
gravitational waves and finally obtain the energy spectrum of GWB
we may probe in future observations. Section V contains discussion
and conclusions.

\section{A review: background and linear perturbations}

As proposed in Ref. \cite{Cai:2007qw}, a model of Quintom Bounce
can be described in terms of scalar fields which minimally couple
to the four dimensional Einstein's gravity. The simplest model is
realized by two scalar fields, with its Lagrangian given by
\begin{eqnarray}\label{lagrangian}
{\cal
L}=-\frac{1}{2}\partial_\mu\phi\partial^\mu\phi+\frac{1}{2}\partial_\mu\psi\partial^\mu\psi-V(\phi,\psi)~,
\end{eqnarray}
in a spatially flat Friedmann-Robertson-Walker (FRW) universe
where the signature of the metric is taken to be $(-,+,+,+)$. Here
the essential component is the scalar field $\psi$. It plays a
crucial role in giving a bouncing solution smoothly. Without it,
the model in (1) will be similar to the traditional inflation with
a single scalar field, which as we know suffers from the problem
of the initial singularity.

In collaboration with Qiu, Xia and Li\cite{Cai:2008qb}, we take
the potential to be only the function of the field $\phi$, and
specifically of Coleman-Weinberg form \cite{Coleman:1973jx}:
\begin{eqnarray}
V=\frac{1}{4}\lambda\phi^4 \left( \ln\frac{|\phi|}{v}-\frac{1}{4}
\right) + \frac{1}{16}\lambda v^4~,
\end{eqnarray}
which takes its maximum value $\lambda v^4/16$ at $\phi=0$ and
vanishes at the minima when $\phi=\pm v$. Therefore, the scalar
field $\psi$ merely affects the evolution around the bounce but
decays out quickly when away from the bouncing point.

\begin{figure}[htbp]
\includegraphics[scale=0.75]{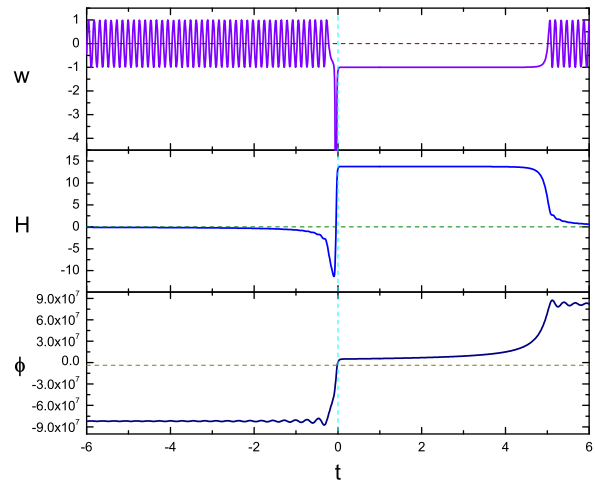}
\caption{A plot of the evolution of the EoS $w$, the hubble
parameter $H$ and the canonical field $\phi$ in the model
(\ref{lagrangian}) where a Coleman-Weinberg potential is
considered. In the numerical calculation we choose the parameters
$\lambda=8.0\times10^{-14},~v=0.82M_{pl}$, and the initial
conditions as:
$\phi=-0.82M_{pl},~\dot\phi=3.0\times10^{-10}M_{pl}^2,~\psi=-0.72M_{pl},~\dot\psi=5.0\times10^{-13}M_{pl}^2$
where $M_{pl}\equiv1/\sqrt{G}$.} \label{fig1:background}
\end{figure}

This model performs an interesting evolution for the background
universe \cite{Cai:2008qb}. Initially, $\phi$ stays at its left
vacuum $-v$ and the kinetic term of $\psi$ is sufficiently small.
This looks very natural and there is no argument for any fields
being outside the Planck scale, so it does not suffer the initial
condition problem which appears in the usual inflationary model
\cite{Brandenberger:1999sw,Linde:2005ht}. At the beginning of the
evolution, the field $\phi$ oscillates around the vacuum point
$-v$, so the EoS of the universe oscillates about $w=0$ and
averagely the state looks similar to a matter dominated one. Since
the universe is contracting, the amplitude of the canonical field
oscillation becomes larger and larger, in the meanwhile the
contribution of $\psi$ also becomes important. When the field
reaches the plateau, the energy density of $\phi$ would be
cancelled by that of $\psi$ and so the bounce happens. After the
bounce, as the field $\phi$ moves forward slowly along the
plateau, the universe enters into a slow-roll phase and the EoS of
the universe is approximately $-1$, very much alike inflation. The
process to link the contraction and expansion is a smooth bounce,
during which the evolution of the hubble parameter can be treated
as a linear function of the cosmic time $t$ approximately.
Finally, when the field $\phi$ ``drops" into the right vacuum
$+v$, it will oscillate again and reheat the universe. To present
the above analysis clearly, we give the numerical calculation of
background evolution in Fig. \ref{fig1:background}. From this
figure, we can see that, although the potential is symmetric
w.r.t. the field $\phi$, the background evolution is asymmetric
w.r.t. the cosmic time.

Note that, the asymmetry of the evolution is a significant character
since we are able to obtain a scale-invariant spectrum in virtue of
this asymmetry. The reason is as follows. Since the evolution of the
universe is asymmetric w.r.t. the bounce point, it is possible for
the primordial fluctuations keeping inside the horizon when the
universe is contracting. And if so, while these perturbations are
able to escape to the super-hubble region during the inflationary
period, the initial Bunch-Davies condition for them can be basically
saved and transferred through the bounce. Therefore, this scenario
provides a possible approach to obtaining a scale-invariant
spectrum. However, as will be shown in the following, we find that
the sub-hubble perturbations still deviate a little from the pure
incoming plane wave on the matching surface between the bouncing
phase and the expanding phase. This deviation would bring some
wiggles on the corresponding scale of the primordial power spectrum,
which has been shown in Ref. \cite{Cai:2008qb} can be detected by
the future observations.

Now we take a brief review of the linear curvature perturbation
(for details we refer to Ref. \cite{Cai:2008qb}, and see Ref.
\cite{MFB} for a comprehensive review), then study its evolution
in the model of Quintom Bounce and obtain the expression for the
primordial power spectrum. Under the longitudinal
(conformal-Newtonian) gauge, the metric perturbations are given by
\begin{eqnarray}
ds^2=a^2(\tau) \left[-(1+2\Phi)d\tau^2+(1-2\Psi)dx^idx^i \right]~,
\end{eqnarray}
where we introduce the comoving time $\tau$ defined by
$d\tau=dt/a$. We start with the equation of motion of the
gravitational potential
\begin{eqnarray}\label{eomp}
\Phi''+2({\cal H}-\frac{\phi''}{\phi'})\Phi'+2({\cal H}'-{\cal
H}\frac{\phi''}{\phi'})\Phi-\nabla^2\Phi \nonumber\\=8\pi G
(2{\cal H}+\frac{\phi''}{\phi'})\psi'\delta\psi~,
\end{eqnarray}
where ${\cal H}\equiv{a'}/{a}$ is the comoving hubble parameter
and the prime denotes the derivative with respect to the comoving
time. This equation can be derived from the basic perturbation
equations directly (we refer the complete derivation to Ref.
\cite{Cai:2007zv}).

As we have pointed out in describing the background evolution, the
energy density of the scalar $\psi$ is usually negligible away
from the bouncing point and hence we have $\psi'\simeq 0$.
Therefore it is a good approximation to neglect the right hand
side of Eq. (\ref{eomp}) when the universe is far away from the
bounce. Moreover, we find that there is the relation $2{\cal
H}+\frac{\phi''}{\phi'} \simeq 0$ in the bouncing phase and so the
perturbation of $\psi$ decouples from Eq. (\ref{eomp}), which has
been checked in Ref. \cite{Cai:2007zv}. Thus, in the paper we will
neglect the r.h.s. of Eq. (\ref{eomp}), and just focus on the
adiabatic fluctuations which can be determined by a single
physical scalar field degree of freedom $\Phi$.

If the evolution of the background is known, all other
perturbation variables can be determined from $\Phi$. A frequently
used variable is the curvature perturbation in comoving
coordinates $\zeta$ which is defined as,
\begin{eqnarray}
\zeta\equiv\Phi+\frac{{\cal H}}{{\cal H}^2-{\cal H}'}(\Phi'+{\cal
H}\Phi)~,
\end{eqnarray}
and this variable can be calculated from $\Phi$ and the background
parameters. For example, when the universe is in a nearly
de-Sitter like expansion, there is a simple relation
$\zeta\simeq\Phi/\epsilon$ with the slow-roll parameter
$\epsilon\equiv-\dot H/H^2$. In usual case this variable is well
known to describe the adiabatic perturbations on large scales
since it is a conserved quantity on super-hubble scales according
to the equation
\begin{eqnarray}
\zeta'=\frac{2k^2(\Phi'+{\cal H}\Phi)}{9 (1+w) {\cal H}^2}~,
\end{eqnarray}
and its dynamics can be simply and conveniently described by the
equation of motion for Mukhanov-Sasaki variable \cite{Mukhanov}.
However, this equation becomes ill-defined both at the bounce
point with a vanishing hubble parameter $H=0$ and the cosmological
constant boundary $w=-1$. This point has been remarked in Ref.
\cite{Cai:2007zv,Brandenberger:2007by,Xia:2007km}. Consequently,
we investigate the evolution of the gravitational potential $\Phi$
directly in deriving the curvature perturbation, and then moves to
$\zeta$ after the universe entering the expanding phase.

As introduced in the above, the universe in this model experiences
three phases, a contracting one, a bounce, and finally an
inflationary one. We can resolve the perturbation equation in each
phase respectively. In the detailed calculation, we take the
Bunch-Davies vacuum as the initial condition $\Phi_k\sim\frac{4\pi
G}{\sqrt{2k^3}}|\dot\phi|e^{-ik\tau}$, and then make those
solutions smoothly pass through the linking point applying the
matching conditions\cite{Hwang:1991an} (see also
\cite{Deruelle:1995kd,Durrer:2002jn,Copeland:2006tn} for a recent
study). The detailed derivation has been presented in Ref.
\cite{Cai:2008qb}, and here we give the dominant part of the final
gravitational potential directly:
\begin{eqnarray}\label{Fk}
\Phi_k
 &\simeq& \frac{4\pi
 G}{\sqrt{2k^3}}|\dot\phi| e^{-ik\tau}
 \nonumber\\
 &\times&
 \left\{1-\frac{3{\cal H}_{B-}e^{ik/{\cal H}_{B+}}}{2k}
 \sin[k/{\cal H}_{B+}] \right\}~,
\end{eqnarray}
where ${\cal H}_{B-}$ and ${\cal H}_{B+}$ represent the comoving
hubble parameter at the beginning and the end of the bouncing
phase respectively. By comparing the coefficient (\ref{Fk}) and
the initial form of $\Phi$, obviously the dominant mode of the
gravitational potential deviate from the Bunch-Davies initial
condition when the inflationary phase takes place. Moreover, we
have the approximate relation $\zeta\simeq\Phi/\epsilon$. So we
eventually have the expression of the primordial power spectrum
for the curvature perturbation\footnote{This result seems similar
to an inflation theory with its initial condition as
$\alpha$-vacuum\cite{Danielsson:2002kx,Easther:2002xe}. However,
these two are different in the detailed characteristics of the
perturbations and the basic mechanisms for the generation of those
perturbations.}
\begin{eqnarray}\label{Pzeta}
P_{\zeta}\simeq\frac{8}{3}G^2\frac{\rho}{\epsilon}\bigg\{
1-\frac{3{\cal H}_{B-}}{2k}\sin\frac{2k}{{\cal H}_{B+}} \bigg\}~.
\end{eqnarray}
From this result, one obviously find that the first term provides
a nearly scale-invariant spectrum which is consistent with current
cosmological observations. However, the second term shows that
there is a wiggle on the spectrum, due to the modified initial
condition by the bounce relative to the standard inflation.
Apparently, this oscillation term could affect the CMB temperature
power spectrum and LSS matter power spectrum\cite{Cai:2008qb}.

\section{Bispectrum and the non-linear parameter $f_{NL}$}

In this section we investigate the evolution of the nonlinear part
of scalar perturbations in our model. As we have discussed in the
last section, the universe experiences a nearly exponential
expansion after the bounce, and so we can deal with the curvature
perturbation of Quintom Bounce as in inflation theory except for
the initial condition being modified. Thus, the slow-roll
parameters can be defined in our model, with $\epsilon\equiv-\dot
H/H^2$ and
$\eta\equiv-\frac{\ddot\phi}{H\phi}+\frac{\dot\phi^2}{2H^2}$.

To expand the action to the third order of $\zeta$ and drop the
terms suppressed by slow-roll parameters $\epsilon$ and $\eta$, we
have the final cubic action for the curvature perturbation during
the inflationary phase\cite{Maldacena:2002vr},
\begin{eqnarray}
S_3=\int dtd^3x \bigg[ g
a^2\dot\zeta^2\partial^{-2}\dot\zeta+2f(\zeta)\frac{\delta{\cal
L}}{\delta\zeta}|_1 \bigg]~,
\end{eqnarray}
where
\begin{eqnarray}
f(\zeta)=\frac{3\epsilon-2\eta}{4}\zeta^2+\frac{\epsilon}{2}\partial^{-2}(\zeta\partial^2\zeta)~.
\end{eqnarray}
and $g=4a^3\epsilon^2H$.

Note that the last term $\frac{\delta{\cal L}}{\delta\zeta}|_1$ in
the third-order action can be absorbed by field redefinitions of
$\zeta$. It is taken as
\begin{eqnarray}
\zeta={\tilde\zeta}+f({\tilde\zeta})~.
\end{eqnarray}
This field redefinitions do not affect any other of the ${\cal
O}(\zeta^3)$ terms in the third-order action, since the term $f$
are quadratic in $\zeta$. Moreover, the last term in the
third-order action can be cancelled by one extra quadratic part of
this field redefinitions which is proportional to the first-order
equations of motion $\frac{\delta L}{\delta \zeta}|_1$ exactly.

After deriving out the third-order action, now we are able to get
the three point function for our model. It can be computed using
the path integral formalism in the interaction picture as follows
\begin{eqnarray}
&&<\zeta(t,{{\bold k}_1})\zeta(t,{{\bold k}_2})\zeta(t,{{\bold
k}_3})> = \nonumber\\ && i\int^t_{t_i}d\tilde t <[\zeta(t,{{\bold
k}_1})\zeta(t,{{\bold k}_2})\zeta(t,{{\bold k}_3}),L_3(\tilde
t)]>,
\end{eqnarray}
where $t_i$ denotes an initial time for the modes deep inside the
horizon, and $L_3$ is the third order perturbative lagrangian.

Recall in Eq. (\ref{Pzeta}) we have obtained the linear curvature
perturbation which deviates from that in the standard inflation
theory by multiplying an extra factor defined as follows,
\begin{eqnarray}
C(k)=1-\frac{3{\cal H}_{B-}}{2k}\sin\frac{2k}{{\cal H}_{B+}}+...~.
\end{eqnarray}
Considering this factor in the calculations, the leading order
contributions to the three point correlator in our model are
listed in the following,\\
(a): contribution from $\dot\zeta^2\partial^{-2}\dot\zeta$. To
make the expression simplified, we define $\bar
P_\zeta=\frac{8}{3}G^2\frac{\rho}{\epsilon}$ which is just the
curvature spectrum in inflation models and $K=\sum_ik_i$. So we
have the three point correlator from
$\dot\zeta^2\partial^{-2}\dot\zeta$:
\begin{eqnarray}
(2\pi)^7\delta^{(3)}(\sum_i{\bold k}_i)(\bar P_\zeta)^2
(\prod_i\frac{C(k_i)}{k_i^3}) \times
\frac{\epsilon}{K}\sum_{i<j}k_i^2k_j^2~.
\end{eqnarray}
(b): contribution from the field redefinition.
\begin{eqnarray}
&(2\pi)^7\delta^{(3)}(\sum_i{\bold k}_i)(\bar P_\zeta)^2
(\prod_i\frac{C(k_i)}{k_i^3})& \nonumber\\ &\times \bigg[
\frac{3\epsilon-2\eta}{8}\sum_i\frac{k_i^3}{C(k_i)}+\frac{\epsilon}{8}\sum_{i\neq
j}\frac{k_ik_j^2}{C(k_i)} \bigg]&
\end{eqnarray}

Moreover, since non-Gaussianity measures the deviation of CMB
power spectrum from the Gaussian distribution, we can define a
non-linear parameter $f_{NL}$ as follows,
\begin{eqnarray}
\zeta=\zeta_g+\frac{3}{5}f_{NL}(\zeta_g^2-<\zeta_g^2>)~.
\end{eqnarray}
So we eventually have $f_{NL}$ to characterize the size of
non-Gaussianity,
\begin{widetext}
\begin{eqnarray}
f_{NL}=\frac{10}{3}\bigg[
\frac{\epsilon}{K}\sum_{i<j}k_i^2k_j^2+\frac{3\epsilon-2\eta}{8}\sum_i\frac{k_i^3}{C(k_i)}+\frac{\epsilon}{8}\sum_{i\neq
j}\frac{k_ik_j^2}{C(k_i)}
\bigg]{\bigg/}(\sum_i\frac{k_i^3}{C(k_i)})~.
\end{eqnarray}
\end{widetext}

\begin{figure}[htbp]
\includegraphics[scale=0.7]{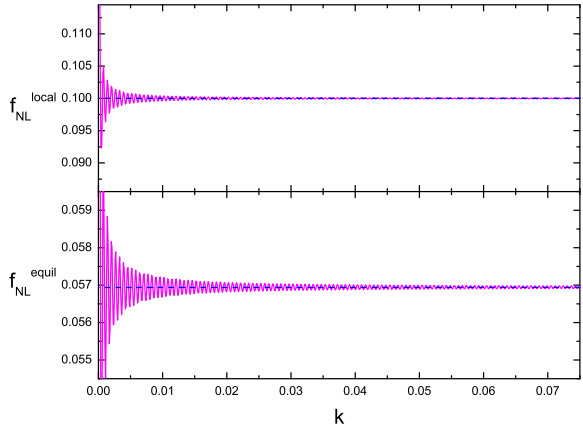}
\caption{The plots of the non-linear parameters $f_{NL}$ with
different forms predicted by our model are shown by the magenta
solid lines. To make the comparison, we also give $f_{NL}$ predicted
by a single scalar inflation model in blue dotted lines.}
\label{fig2:fnl}
\end{figure}

There are two limiting cases of non-Gaussianity, which are of
particular interests for observations. These are equilateral form
($k_1\sim k_2\sim k_3$) and local form ($k_1\sim k_2\gg k_3$). To
the case of equilateral form, $f_{NL}$ is given by
\begin{eqnarray}
f_{NL}^{equil}\simeq\frac{10}{9}\bigg[
\bigg(\frac{15}{8}+C(k)\bigg)\epsilon-\frac{3}{4}\eta \bigg]~,
\end{eqnarray}
where we have taken the limit $k=k_1=k_2=k_3$. For the local form,
which corresponds to that $k_3$ mode exits horizon much earlier
than the other two, we have the non-linear parameter
\begin{eqnarray}
f_{NL}^{local}\simeq \frac{5}{3}\bigg[ \bigg(2+C(k)\bigg)\epsilon
-\eta \bigg]~,
\end{eqnarray}
where the limit is taken as $k=k_1=k_2\gg k_3$. One may notice
that, the above results reduce to the single scalar slow-roll
inflationary model when $C(k)\rightarrow 1$. However, since the
factor $C(k)$ has modified the initial condition of curvature
perturbation when the universe enters inflationary stage, it can
bring an oscillation signature on the size of non-Gaussianity as
well. In order to compare our result with the non-Gaussianity
predicted by usual inflation model, we plot $f_{NL}$ of
equilateral and local forms in Fig. \ref{fig2:fnl}.

Our results show that the non-Gaussianity in the Quintom Bounce
model is still suppressed by the slow-roll parameters. However,
there is an oscillation signature on $f_{NL}$ and the maximal
value of $f_{NL}$ is bigger than that in single scalar slow-roll
inflationary models. The reason for this effect is that the
dominant modes of the curvature perturbations have deviated from
the Bunch-Davies form when they pass through the bounce and enter
the inflationary stage. This is similar to the cases with
non-Gaussianity generated from a modified initial condition, for
example see Refs. \cite{Chen:2006nt,Holman:2007na,Chen:2007gd}.

\section{Gravitational wave background}

Now we turn to consider the evolution of gravitational wave
background from the tensor part of the primordial metric
perturbations. In order to standardize the derivation, we use the
same convention as in Ref. \cite{Cai:2007xr}. To begin with, we
give the metric containing the tensor perturbations in the flat
FRW background as follows,
\begin{eqnarray}
ds^2=a(\tau)^2[-d\tau^2+(\delta_{ij}+\bar h_{ij})dx^idx^j]~,
\end{eqnarray}
where the Latin indexes represent spatial coordinates. Here the
tensor perturbation $\bar h_{ij}$ satisfies the following
constraints:
\begin{eqnarray}
\bar h_{ij}=\bar h_{ji}~;~~\bar h_{ii}=0~;~~\bar h_{ij,j}=0~.
\end{eqnarray}
Due to these constraints, we only have two degrees of freedom in
$\bar h_{ij}$ which correspond to two polarizations of
gravitational waves.

By adding the anisotropic part of the stress tensor $\sigma_{ij}$,
we have the equation of motion for tensor perturbations,
\begin{eqnarray}
\bar h_{ij}''+2\frac{a'}{a}\bar h_{ij}'-\nabla^2\bar h_{ij}=16\pi
Ga^2\sigma_{ij}~.
\end{eqnarray}
The Fourier transformations of the tensor perturbations and
anisotropic stress tensor are give by,
\begin{eqnarray}
\bar h_{ij}(\tau, {\bf x})=\sqrt{16\pi
G}\int\frac{d^3k}{(2\pi)^\frac{3}{2}}H_{ij}(\tau, {\bf k})e^{i{\bf
k}{\bf x}}~,\\
\sigma_{ij}(\tau, {\bf x})=\sqrt{16\pi
G}\int\frac{d^3k}{(2\pi)^\frac{3}{2}}\Sigma_{ij}(\tau, {\bf
k})e^{i{\bf k}{\bf x}}~.
\end{eqnarray}

Note that, what we are usually interested in are the distribution
of the spectra of gravitational waves and the corresponding
spectral index. Based on the above formalism, the tensor power
spectrum can be written as,
\begin{eqnarray}\label{tensor power}
P_T(k,\tau)&\equiv\frac{d\langle0|\bar{h}_{ij}^{2}|0\rangle}{d\,{\rm
ln}\,k}=32\pi G\frac{k^3}{(2\pi)^2}|H_{ij}(\tau, {\bf k})|^2~,
\end{eqnarray}
and the definition of tensor spectral index $n_T$ is given by
\begin{eqnarray}
n_T\equiv\frac{d\,{\rm ln}\,P_T}{d\,{\rm ln}\,k}~.
\end{eqnarray}
The GWB we observed today is characterized by the energy spectrum,
\begin{eqnarray}\label{tensor energy}
\Omega_{GW}(k,
\tau)\equiv\frac{1}{\rho_c(\tau)}\frac{d\langle0|\rho_{GW}(\tau)|0\rangle}{d\,{\rm
ln}\,k}~,
\end{eqnarray}
where $\rho_{GW}(\tau)$ is the energy density of gravitational
waves, and the parameter $\rho_c(\tau)$ is the critical density of
the universe. In respect that the GWB we observed has already
reentered the horizon, the modes should oscillate in the form of a
sinusoidal function. Consequently, we can make use of the
Friedmann equation $H^2(\tau)=\frac{8\pi G}{3}\rho_c(\tau)$ and
then deduce the relation between the power spectrum and the energy
spectrum as follows,
\begin{eqnarray}\label{approximate tensor energy}
\Omega_{GW}(k,\tau)\simeq\frac{1}{12}\frac{k^2}{a^2(\tau)H^2(\tau)}P_T(k,
\tau)~,
\end{eqnarray}
which will be used in the following calculations.

\subsection{Tensor perturbations}

\begin{figure}[htbp]
\includegraphics[scale=0.35]{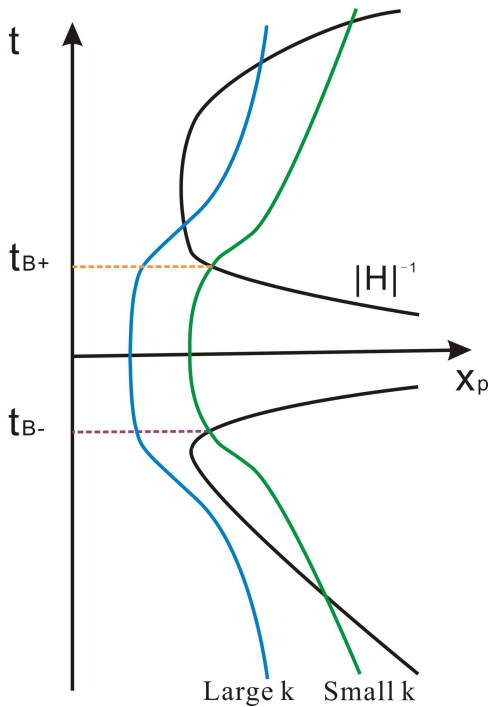}
\caption{A sketch plot of the evolution of tensor perturbations
with different comoving wave number $k$ in our model.}
\label{fig3:sketch}
\end{figure}

Now we follow one Fourier mode of the tensor perturbations,
labelled by its comoving wave number $k$, and find that there are
two paths which are different in the times of crossing the hubble
radius. The evolution of tensor perturbations is sketched in Fig.
\ref{fig3:sketch}. Initially all the perturbations stay inside the
hubble radius in the far past. Since the hubble radius shrinks in
the contracting phase, those modes with small comoving wave number
exit the hubble radius while the large $k$ scales still keep
inside. When the bounce takes place, all the perturbations will
keep inside the hubble radius because at that moment the hubble
radius diverges. After that the bounce is followed by a slow-roll
expanding phase, so these Fourier modes will escape out if the
efolds for the post-bounce inflationary period is large enough.
After that, these modes will reenter the hubble radius at late
times after the slow-roll expanding phase has finished.

Therefore, we can classify the tensor perturbations with different
comoving wave numbers $k$ to two categories, as the two lines
sketched in Fig. \ref{fig3:sketch}. The blue line denotes a mode
with a scale $k$ which is large enough to keep it inside the
hubble radius through the contracting phase and the bounce, and
then escape outside during the post-bounce inflationary period;
the green line consists of a mode with small $k$ so that it exits
the hubble radius in the contracting phase, then is pushed inside
during the bounce and soon is pulled outside before the slow-roll
expanding phase happens.

Due to the symmetry of $H_{ij}$, we can express the two
polarizations as one function
$v\equiv\frac{a}{\sqrt{2}}(H_{11}+iH_{22})$. Neglecting the
anisotropic stress tensor in the very early universe, we obtain
the equation of motion for $v$:
\begin{eqnarray}\label{eomt}
v''+k^2v-\frac{a''}{a}v=0~.
\end{eqnarray}

Following the background evolution of the universe, we obtain
three solutions of gravitational waves similar to what we did with
scalar perturbations. For the universe which is contracting with
its EoS oscillating around $w=0$, we have
\begin{eqnarray}
v&=&(\tau-\tilde\tau_{B-})^{\frac{1}{2}} \bigg\{
A^T_kH_{\frac{3}{2}}^{(1)}[k(\tau-\tilde\tau_{B-})]
\nonumber\\&&+B^T_kH_{\frac{3}{2}}^{(2)}[k(\tau-\tilde\tau_{B-})]
\bigg\}~,
\end{eqnarray}
where $\tilde\tau_{B-}=\tau_{B-}+2/{{\cal H}_{B-}}$. Here
$H^{(1)}_\nu$ and $H^{(2)}_\nu$ are the $\nu$-th Hankel function
of the first kind and second kind respectively. Besides, the
parameters $A^T_k$ and $B^T_k$ can be determined by the initial
condition for gravitational waves, which is usually taken as
Bunch-Davies vacuum $v\sim e^{-ik\tau}/{\sqrt{2k}}$. So we have
$A^T_k=0$ and $B^T_k=-{\sqrt{\pi}}/2$. Therefore, the asymptotic
forms of the solution to the tensor perturbation in the
contracting phase is
\begin{eqnarray}
  v(k,\tau)= \left\{ \begin{array}{c}
    -i\frac{k^{\frac{3}{2}}}{8\sqrt{2}}(\tau-\tilde\tau_{B-})^2,~~{\rm outside~horizon}; \\
    \\
    \frac{1}{\sqrt{2k}}e^{-ik(\tau-\tilde\tau_{B-})},~~{\rm
    inside~
    horizon}.
\end{array} \right.  \label{v1}
\end{eqnarray}

When the universe undergoes the bouncing phase, we have the
approximate relation that $\frac{a''}{a}\simeq\frac{4}{\pi}\alpha
a_B^2=\frac{y}{2}$. To solve Eq. (\ref{eomt}), we have
\begin{eqnarray}
  &&v(k,\tau)= \nonumber\\ &&\left\{ \begin{array}{c}
    C^T_k\cos[l(\tau-\tau_B)]+D^T_k\sin[l(\tau-\tau_B)],~~{\rm k^2\geq\frac{y}{2}}; \\
    \\
    C^T_ke^{l(\tau-\tau_B)}+D^T_ke^{-l(\tau-\tau_B)},~~{\rm
    k^2<\frac{y}{2}},
\end{array} \right.  \label{v2}
\end{eqnarray}
where we define $l^2=|k^2-\frac{y}{2}|$. Since the hubble
parameter approaches zero when the universe is bouncing from a
contraction to an expanding phase, all the modes of the
perturbations would return to the sub-hubble region. However, from
the above solution we interestingly find that, $k_{ph}^2(\sim
k^2/a_B^2)$ and $\dot H(\sim\alpha)$ are comparable.

After the bounce, the slow-roll expanding phase takes place which
drives the universe to inflate like a de-Sitter spacetime. In this
case, the solution to the gravitational waves is given by
\begin{eqnarray}
v&=&(\tau-\tilde\tau_{B+})^{\frac{1}{2}} \times \bigg\{
E^T_kH_{\nu}^{(1)}[k(\tau-\tilde\tau_{B+})] \nonumber\\
&&+F^T_kH_{\nu}^{(2)}[k(\tau-\tilde\tau_{B+})] \bigg\}~,\label{v3}
\end{eqnarray}
where $\nu=\frac{1}{2}+\frac{1}{1-\epsilon}\simeq\frac{3}{2}$.
This solution has an asymptotic form,
\begin{eqnarray}\label{v3a}
v\simeq-i\sqrt{\frac{2}{\pi}}k^{-\frac{3}{2}}(\tau-\tilde\tau_{B+})^{-1}(E^T_k-F^T_k)~,
\end{eqnarray}
after the modes exit the horizon.

Having obtained the solutions of the tensor perturbations in
different phases, now we need to match these solutions and
determine the coefficients $C^T_k$, $D^T_k$, $E^T_k$ and $F^T_k$
respectively. This procedure is much similar to the matching
process of scalar perturbations as done in the previous section.
For a non-singular bounce scenario such as the Quintom Bounce
model, the continuity of background evolution implies that both
$v$ and $v'$ are able to pass through the bounce smoothly. So we
match $v$ and $v'$ in (\ref{v1}) and (\ref{v2}) on the surface
$\tau_{B-}$, and those in (\ref{v2}) and (\ref{v3}) on the surface
$\tau_{B+}$. With these matching conditions, we can determine all
the coefficients and finally get $E^T_k-F^T_k$.

However, as what we have analyzed at the beginning of this
section, there are two paths for the tensor perturbations to
evolve from a contracting phase to an expanding phase. So there
are two possible results for $E^T_k-F^T_k$. For the first case,
the comoving wave number is large enough so that the tensor
perturbations have never escape outside the hubble radius, thus we
have
\begin{eqnarray}\label{EF1}
|E^T_k-F^T_k|\simeq\frac{\sqrt{\pi}}{2}|1-\frac{\sigma}{4}e^{i\frac{2k}{{\cal
H}_{B+}}}(1-e^{2ik\delta\tau_B})|~,
\end{eqnarray}
where $\sigma\equiv\frac{y}{2k^2}$, and
$\delta\tau_B=\tau_{B+}-\tau_{B-}$. For the second case where the
modes of gravitational waves are in small $k$ region, the
expression $E^T_k-F^T_k$ is given by
\begin{eqnarray}\label{EF2}
E^T_k-F^T_k\simeq-\frac{\sqrt{\pi}}{8}\frac{{\cal H}_{B-}(2l+{\cal
H}_{B-})}{{\cal H}_{B+}(l-{\cal H}_{B+})}e^{-l\delta\tau_B}~.
\end{eqnarray}

\begin{figure}[htbp]
\includegraphics[scale=0.75]{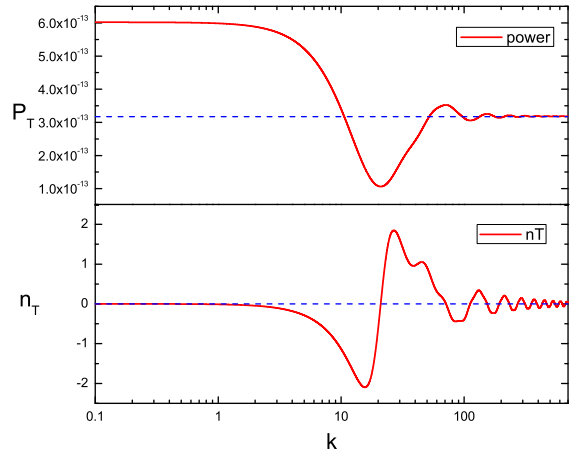}
\caption{The red solid curve represents the primordial power
spectrum $P_T$ and the spectral index $n_T$ of tensor perturbations
in our model. The blue dash curve give the primordial power spectrum
and the spectral index in a single scalar inflation model. In the
figure, we take the values of parameters the same as in Fig.
\ref{fig1:background}.} \label{fig4:pnt}
\end{figure}

Based on the above analysis, now we are able to derive the
primordial power spectrum of gravitational waves. From the
definition of Eq. (\ref{tensor power}), the primordial power
spectrum is given by
\begin{eqnarray}
P_T(k)=\frac{64GH^2}{\pi^2}|E^T_k-F^T_k|^2~.
\end{eqnarray}
From Eqs. (\ref{EF1}) and (\ref{EF2}), we can read that the
spectrum are scale-invariant both at the large $k$ and small $k$
region, but oscillate when $k$ is near to a critical value
$\sqrt{\frac{y}{2}}$. To illustrate the above analysis clearly, we
do the numerical calculation and plot the results of primordial
tensor power spectrum and the corresponding spectral index in Fig.
\ref{fig4:pnt}.

One can see that in Fig. \ref{fig4:pnt}, when the value of $k$ is
large enough, the red solid lines converge at the blue dash lines
with an oscillation. The amplitude of the oscillation gets the
largest value when $k$ approaches the neighborhood of the critical
value $\sqrt{\frac{y}{2}}$, and soon drop down to a minimal value
when $k$ gets smaller. This damping effect is caused by the
modified dispersion relation of the tensor perturbations when they
pass through the bouncing phase\footnote{A similar scenario of the
primordial gravitational waves has been considered in Ref.
\cite{Cai:2007xr}, where the authors have considered the damping
effects from the spacetime
noncommutativity. 
}. However, when the comoving wave number $k$ gets even smaller,
the power spectrum is able to climb up and finally reaches a
certain value with its spectral index returning to zero again.

\subsection{Energy Spectrum of Today's GWB}

In the above section we discussed the behavior of tensor
perturbations exhibited in primordial power spectrum and the
spectral index. However, we are more interested in how to
recognize these perturbations in the GWB nowadays. Since the
primordial gravitational waves are distributed in every frequency,
once the effective co-moving wave number is less than $aH$, the
corresponding mode of gravitational waves would escape the horizon
and be frozen until it reenters the horizon. The relation between
the time when tensor perturbations exit the horizon and the time
when they return is $a_{out}H_{out}=a_{in}H_{in}$. Therefore, we
have the conclusion that, the earlier the perturbations escape the
horizon, the later they re-enter it. Moreover, once the effective
co-moving wave number is larger than $aH$, the perturbations begin
to oscillate like the plane wave, as shown in Fig.
\ref{fig3:sketch}.

To relate the power spectrum observed today to the primordial one,
one can define a transfer function $T(k,\tau)$, given by Refs.
\cite{Boyle:2005se,Cai:2007xr}:
\begin{eqnarray}
T(k,\tau)\simeq\frac{0.80313}{2\pi}\bigg(\frac{1+z(\tau)}{1+z_k}\bigg)^2\Gamma^2(\alpha+\frac{1}{2})(\frac{2}{\alpha})^{2\alpha}~,
\end{eqnarray}
where $\alpha=\frac{2}{1+3w}$ is determined by the EoS of the
universe, $z(\tau)$ is the redshift at the moment $\tau$ and $z_k$
is the redshift when the $k$ mode of gravitational wave reenters
the horizon. Here the factor $0.80313$ comes from the damping
effect of freely streaming neutrinos \cite{Weinberg:2003ur}.
Moreover, the factor $(\frac{1+z(\tau)}{1+z_k})^2$ describes the
redshift-suppressing effect on the primordial gravitational waves.
The rest factor shows that, when the gravitational waves reenter
the horizon, there is a ``wall" lying on the horizon which affects
the tensor spectrum.

Considering today our universe is dominated by dark energy of
which the EoS is $w\simeq-1$, we are able to obtain today's
transfer function. Then we can get today's tensor power spectrum
\begin{eqnarray}
P_T(k,\tau_0) = \frac{0.80313}{2(1+z_k)^2}
\frac{64GH_i^2}{\pi^2}|E_k^T-F_k^T|^2~,
\end{eqnarray}
where $H_i$ represents the hubble parameter in the inflationary
stage. Eventually, the present energy spectrum of GWB is given by
$\Omega_{GW}(k,\tau_0)=\frac{1}{12}\frac{k^2}{(a_0H_0)^2}P_T(k,\tau_0)$.

\begin{figure}[htbp]
\includegraphics[scale=0.8]{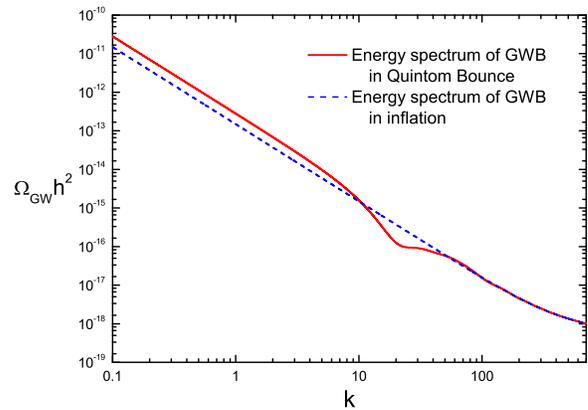}
\caption{The red solid curve represents the energy spectrum
$\Omega_{GW}$ in our model. The blue dash curve gives the energy
spectrum in a single scalar inflation model. In the figure, we take
the values of parameters the same as in Fig. \ref{fig1:background}.}
\label{fig5:omega}
\end{figure}

In Fig. \ref{fig5:omega} we plot the numerical results of the
energy spectrum in our model. One can see that, when the frequency
of GWB is large enough, the tensor energy spectrum of our model
would agree with the prediction of the single scalar inflationary
model. However, when the frequency becomes smaller, the physics of
a bounce begins to affect the behavior of the GWB. Moreover, there
is an interesting sunken area in the middle band. This sunken
signal is resulted from the modified dispersion relation of the
tensor perturbations when they pass through the bouncing phase.

\section{Discussion and Conclusions}

Bouncing cosmology, due to the avoidance of the initial
singularity, has attracted a lot of interests in the literature
\cite{Peter,Veneziano,Wands,Piao:2004me,Senatore,Buchbinder:2007ad,ArmendarizPicon:2003qk}
(and see Ref. \cite{Novello:2008ra} for a recent review). However,
since it happens in extremely high energy regime, we hardly
observe a bounce by experiments directly. So it is a debate
whether a bounce has taken place or not. To find the evidences of
a bounce, we need to know what can a bounce leave for
observations. This question is still discussed drastically in the
literature, and one potential clue is to study the primordial
gravitational fluctuations. In the context of the Pre-Big-Bang
scenario and in the cyclic/Ekpyrotic cosmology, the primordial
curvature perturbation strongly depends on the physics at the
epoch of thermalization, and thus an uncertainty of a thermalized
surface is involved \cite{BGGMV,Lyth,Hwang2,Fabio,Tolley:2003nx}.
In the frame of loop quantum cosmology, it is argued that
fluctuations before and after the bounce are largely independent
\cite{Bojowald:2007zza} (yet see Ref. \cite{Corichi:2007am} for
some criticisms). We in this paper have studied the perturbation
theory of a Quintom Bounce model detailedly and show that there
are some imprints of the bounce on CMB observations at large
scales. In the main content we have analyzed both the linear and
non-linear evolutions of scalar modes, and the tensor
perturbations are also considered.

The model we considered is constructed by a double field Quintom
model with a Coleman-Weinberg potential. We firstly have reviewed
the background dynamics of this model, and obtained a
scale-invariant scalar spectrum in virtue of an asymmetry of the
background evolution around the bounce point. A similar but more
phenomenological scenario has been studied in Refs.
\cite{Piao:2003zm} as a possible solution to the suppressed low
multi-poles of the CMB anisotropies. Moreover, since the
gravitational perturbations in sub-hubble region would change
their propagations when pass through the surface between the
contracting phase and the bounce, there would be an oscillation
signature generated both on the linear scalar modes and non-linear
ones. We have also calculated the non-Gaussianity and shown that
the maximal value of non-linear parameter $f_{NL}$ predicted by
our model is mildly bigger than the usual one in single scalar
slow-roll inflation, but the central value is still suppressed by
the slow-roll parameters. So we expect that a large
non-Gaussianity might be generated by other
mechanisms\cite{Li:2008fm} in the frame of Quintom Bounce.

We in the last part of this paper focus on the behavior of the
gravitational waves in Quintom Bounce. Due to the effects of a
bounce, the solution of the tensor perturbation is quite different
from the usual one. In our analysis, we find that the physics of a
bounce would affect the evolution of primordial tensor
perturbations at large scales of the universe, which corresponds
to the physics in very early time. The behavior of the energy
spectrum of the GWB in our model is similar to that in the single
scalar inflationary model in high-frequency regime. However, for
low-frequency regime the difference becomes larger. Moreover,
there is a sunken area in the middle band which links
high-frequency regime and low-frequency regime. If these signals
would be detected, these might act as a smoking gun to the
bouncing cosmology.

\begin{acknowledgments}

We thank Hong Li, Mingzhe Li, Jie Liu, Yun-Song Piao, Taotao Qiu and
Jun-Qing Xia for helpful discussions. This work is supported in part
by National Natural Science Foundation of China under Grant Nos.
10533010 and 10675136 and by the Chinese Academy of Science under
Grant No. KJCX3-SYW-N2.

\end{acknowledgments}

\vfill


\end{document}